\documentclass[article]{aastex631}

\usepackage{amsmath}
\usepackage{rotating}
\newcommand{\kms}{km s$^{-1}$}
\makeatletter
\usepackage[flushleft]{threeparttable}
\usepackage{threeparttable}

\DeclareRobustCommand{\O3}{%
  \mbox{O\check@mathfonts\fontsize\sf@size\z@\selectfont III}%
}
\makeatother

\usepackage{hyperref}
\usepackage{multirow}

\newcommand{\degree}{{}$^\circ$}
\newcommand{\HI}{\ion{H}{1}}

\shorttitle{ALMA View of Positive Black Hole Feedback in the He 2--10}
\shortauthors{Gim \& Reines}

\begin{document}

\title{The ALMA View of Positive Black Hole Feedback in the Dwarf Galaxy Henize~2--10}

\author[0000-0003-1436-7658]{Hansung B. Gim}
\affiliation{Department of Physics, Montana State University, Bozeman, MT 59717, USA}

\author[0000-0001-7158-614X]{Amy E. Reines}
\affiliation{Department of Physics, Montana State University, Bozeman, MT 59717, USA}
\affiliation{eXtreme Gravity Institute, Department of Physics, Montana State University, Bozeman, MT 59717, USA}

\correspondingauthor{Hansung B. Gim}
\email{hansung.gim@montana.edu}

\begin{abstract}

Henize 2--10 is a dwarf starburst galaxy hosting a $\sim10^{6}~M_{\odot}$ black hole (BH) that is driving an ionized outflow and triggering star formation within the central $\sim100$~pc of the galaxy. Here we present ALMA continuum observations from 99 to 340~GHz, as well as spectral line observations of the molecules CO~(1--0, 3--2), HCN~(1--0, 3--2), and HCO$+$~(1--0, 3--2), with a focus on the BH and its vicinity. Incorporating cm-wave radio measurements from the literature, we show that the spectral energy distribution of the BH is dominated by synchrotron emission from 1.4 to~340 GHz with a spectral index of $\alpha\approx-0.5$. We analyze the spectral line data and identify an elongated molecular gas structure around the BH with a velocity distinct from the surrounding regions. The physical extent of this molecular gas structure is $\approx130~{\rm pc}\times30$~pc and the molecular gas mass is $\sim10^{6}~M_{\odot}$. Despite an abundance of molecular gas in this general region, the position of the BH is significantly offset from the peak intensity, which may explain why the BH is radiating at a very low Eddington ratio. Our analysis of the spatially-resolved line ratio between CO J=3--2 and J=1--0 implies that the CO gas in the vicinity of the BH is highly excited, particularly at the interface between the BH outflow and the regions of triggered star formation. This suggests that the cold molecular gas is being shocked by the bipolar outflow from the BH, supporting the case for positive BH feedback.

\end{abstract}


\section{Introduction} \label{sec:intro}

Galaxy evolution is governed by the balance between star formation and quenching \citep{Somerville2015}, and Active Galactic Nuclei (AGN) are thought to play a crucial role in the extinguishing of star formation in massive galaxies. Prior to the last decade, massive black holes (BHs) that sometimes accrete and shine as AGNs were almost exclusively found in giant galaxies \citep[e.g.,][]{kormendy2013}. However, we now know that AGNs in dwarf galaxies are much more common than previously thought and the BHs have typical masses of $M_{\rm BH} < 10^6~M_\odot$ \citep[for reviews, see][]{Greene2020,Reines2022}. Searching for and studying these systems is important for our understanding of the role of AGN fueling and feedback in the low-mass regime and may also provide clues about the origin of the first BH ``seeds" in the early Universe \citep[e.g.,][]{Volonteri2021}.

\begin{figure}
    \centering
    \includegraphics[width=6in]{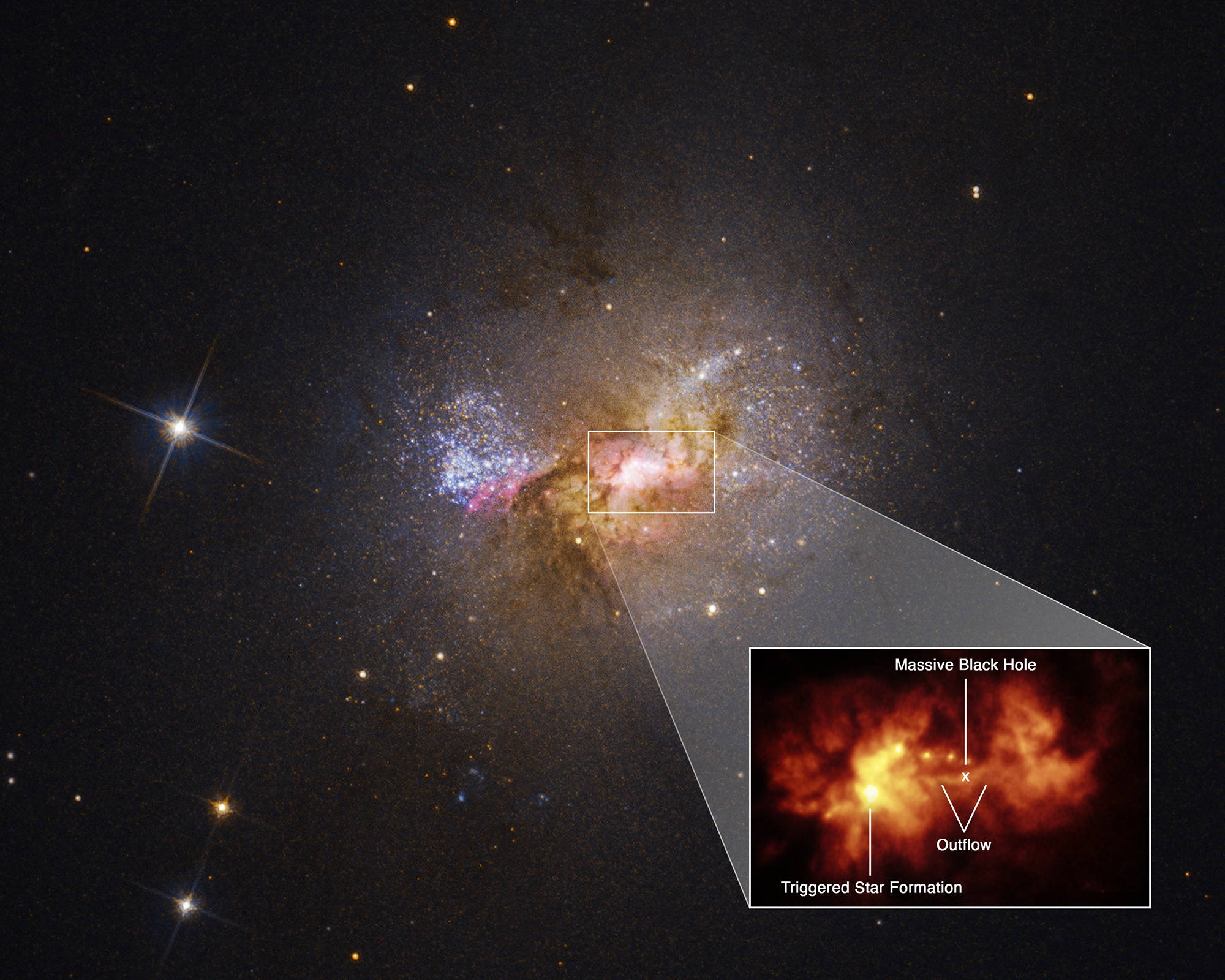}
    \caption{The four color composite image of He 2--10 is generated from F330W (B), F606W (V), F658N (H$_{\alpha}$), and F814W (I) from the \textit{HST}. The white box shows the central 6\arcsec$\times$4\arcsec\ region, and the inset shows the H$\alpha$$+$continuum image presenting the outflow from the BH. (Credit to NASA, ESA, Zachary Schutte, \& Amy Reines) }
    \label{fig:He210_survey}
\end{figure}

Henize 2--10 (He 2--10) is a nearby (d$=$9 Mpc) dwarf starburst galaxy \citep[e.g.,][]{Nguyen2014} that is particularly remarkable for displaying evidence of positive AGN feedback \citep[i.e., enhancing star formation;][see Figure \ref{fig:He210_survey} here]{Schutte2022}. Multiwavelength studies have shown that this galaxy hosts a massive BH in its center \citep[][]{Reines2011} through coincident peaks at radio and X-ray wavelengths \citep{Reines2016} detected with the Very Large Array (VLA) and the {\it Chandra X-ray Observatory} (Chandra), respectively. The BH is also detected as a strong non-thermal source by high-angular resolution observations using the Long Baseline Array \citep[LBA, ][]{Reines2012}. The mass of the central BH is estimated to be $\rm M_{BH} \sim 10^6 \; M_\odot$ based on stellar velocity measurements \citep{Riffel2020} and the scaling relation between total stellar mass and BH mass \citep[][]{Reines2015}. Combined with the X-ray luminosity, this indicates the BH is accreting at a very low Eddington ratio \citep[][]{Reines2016}. A recent study using {\it Hubble Space Telescope} (HST) spectroscopy revealed that the central BH is driving an outflow of ionized gas, which is triggering the formation of star clusters located $\sim$70~pc ($\sim$ 1\farcs5) from the BH \citep[][]{Schutte2022}. This is the first case of positive BH feedback found in a dwarf galaxy and may represent a low-power analog of radio-loud AGNs experiencing ``jet-mode" feedback \citep[][]{Schutte2022}.

Molecular gas and dust can both feed and obscure AGNs. While the cold gas distributions of He 2--10 have been extensively studied for the galaxy as a whole and with a focus on the star-forming regions, very little is known about the molecular gas in the immediate vicinity of the BH. Here we present and analyze high-resolution Atacama Large Millimeter/submillimeter Array (ALMA)\footnote{The National Radio Astronomy Observatory is a facility of the National Science Foundation operated under cooperative agreement by Associated Universities, Inc.} observations of He 2--10 with a specific focus on the central massive BH and its immediate surroundings ($\sim$ 130~pc). A summary of previous studies of the molecular gas in He 2--10 is given below. 

\citet{Baas1994} first observed the three CO transitions (J=1--0, J=2--1, and J=3--2) of He 2--10, showing that hot and cold molecular gas is distributed throughout the galaxy. The first interferometric observations of CO J=1--0 with a synthesized beam size of 6\farcs5$\times$5\farcs5 (284~pc$\times$240~pc) revealed an elongated morphology, indicating that He 2--10 could be a post-merger system and/or experiencing tidal interactions \citep{Kobulnicky1995}. \citet{Meier2001} observed the CO J=3--2 line to investigate the interstellar medium (ISM) conditions in the central 13\arcsec\ (567~pc) and found a gas temperature of $\rm T _{K} \sim 5-10$~K and a number density of $\rm n(H_{2})> 10^{3.5}$~cm$^{-3}$. Higher transitions of CO lines, J=3--2, 4--3, 6--5, and 7--6, were observed by \citet{Bayet2004}. They found that the central region hosts warm and dense molecular gas with T$_{K} \approx 50-100$~K and $\rm n(H_{2}) > 10^{4}$~cm$^{-3}$, which is an order of magnitude higher than the temperature calculated with lower transitions of CO lines (J = 1--0, 2--1 and 3--2) by \citet{Meier2001}. 
\citet{Beck2018} observed the CO J=3--2 line with a synthesized beam size of 0\farcs3 (13 pc) using ALMA and found several clumpy filaments separated spatially and kinematically. Their kinematic analysis led to the conclusion that the molecular filaments feed the star formation in the star clusters. 
\citet{Imara2019} traced the CO J=1--0 line with a synthesized beam size of 0\farcs67$\times$0\farcs58 (29~pc$\times$25~pc) using ALMA and showed that He 2--10 has a more complex morphology with clumps, ring-shaped structures, and high-density peak emissions.

Higher critical density molecular gas tracers have also been observed in He 2--10. 
\citet{Imanishi2007} observed the HCN J=1--0 and HCO$^{+}$ J = 1--0 line with a beam size of 10\farcs8$\times$5\farcs5 (471~pc$\times$240~pc) and reported the ratio of HCN (1--0)/HCO$^{+}$ (1--0) $< 0.6$, which is low compared to Luminous InfraRed Galaxies (LIRGs). The strong HCN J=1--0 line was also recovered by the single dish observations of \citet{Santangelo2009}. \citet{Johnson2018} used ALMA observations tracing HCN J=1--0, HCO$^{+}$ J=1--0, HNC J=1--0, and CCH J=1--0 with a synthesized beam size of 1\farcs7$\times$1\farcs6 (74~pc$\times$70~pc) and showed that HCO$^{+}$ J=1--0 correlates with the thermal radio continuum in the intense central star forming regions. 

Here we present the first study of the molecular gas in the immediate vicinity of the central BH using high-resolution observations with ALMA. We describe the observations and data reduction in \S~\ref{sec:obs}, the analysis and results in \S~\ref{sec:results}, and our conclusions in \S~\ref{sec:conclusions}.

\section{ALMA Observations and Data Reduction} \label{sec:obs}

He 2–10 has been observed by ALMA in different bands for six different projects. The specifics of the ALMA observations are summarized in Table~\ref{tab:alma_obs}, encompassing essential information such as project codes, bands, frequency ranges, spectral lines, total integration time, bandpass, phase, and amplitude calibrators. The names of PIs are presented in the caption located below the table. We downloaded all the data listed in Table~\ref{tab:alma_obs} from the ALMA archive website\footnote{The ALMA archival website is at \url{https://almascience.nrao.edu/aq/}}. The calibrated data were processed using the \textit{scriptForPI.py} with the appropriate version of the Common Astronomy Software Application \citep[CASA, ][]{CASA2022} as outlined in the README files. We also flagged some bad antennas, which improved the sensitivity.

\begin{deluxetable*}{cccccccc}
\tablenum{1}
\tablecaption{ALMA observations of He 2--10 \label{tab:alma_obs}}
\tablewidth{0pt}
\tablehead{\colhead{(1)} & \colhead{(2)} & \colhead{(3)} & \colhead{(4)} & \colhead{(5)} & \colhead{(6)} & \colhead{(7)} & \colhead{(8)} \\ 
\colhead{Project code} & \colhead{Band} & \colhead{Frequency} & \colhead{Lines} & \colhead{Integration} & \colhead{Bandpass$^{a}$} & \colhead{Phase$^{b}$} & \colhead{Amplitude$^{c}$}\\ 
\colhead{} & \colhead{}  & \colhead{(GHz)} &\colhead{}  &\colhead{}  &\colhead{}  &\colhead{}  &\colhead{}  \\ } 
\startdata
\multirow{2}{*}{2011.0.00348.S $^{(1)}$} & \multirow{2}{*}{3} & \multirow{2}{*}{86.95 -- 102.47}  & HCN (1--0) & \multirow{2}{*}{204m 37s} &  \multirow{2}{*}{J053851-440507} & \multirow{2}{*}{J082601-223027} & Titan\\
 & & & HCO$^{+}$ (1--0) &  & & & Ceres \\
\multirow{3}{*}{2012.1.00413.S $^{(2)}$} & \multirow{3}{*}{6} & \multirow{3}{*}{248.79 -- 267.73} & \multirow{2}{*}{HCN (3--2)} & \multirow{3}{*}{286m 47s} & J0750+1231 & \multirow{2}{*}{J0747-3310} & \multirow{2}{*}{J1037-295} \\
 & & & \multirow{2}{*}{HCO$^{+}$ (3--2)} & & J0538-4405 & \multirow{2}{*}{J0826-2230} & \multirow{2}{*}{J0538-4405} \\
 & & & & & J1058+0133 & & \\
2015.1.01569.S $^{(3)}$ & 3 & 111.99 -- 115.18 & \multirow{2}{*}{CO (1--0)} & \multirow{2}{*}{187m 50s} & \multirow{2}{*}{J1037-2934} & J0826-2230 & J1107-4449 \\
2016.1.00027.S $^{(3)}$ & 3 & 99.72 -- 115.17  &  &  & & J0538-4405 & J0846-2607 \\
2016.1.00492.S $^{(4)}$ & 7 & 332.06 -- 348.00  & CO (3--2) & 50m 29s &  J0538-4405 & J0846-2607 & J0538-4405 \\
\multirow{2}{*}{2019.1.01641.S $^{(5)}$} & \multirow{2}{*}{3} & \multirow{2}{*}{96.91 -- 112.51} & \multirow{2}{*}{--} & \multirow{2}{*}{103m 43s} & J0725-0054 & \multirow{2}{*}{J0846-2607} & J0725-0054 \\
 & & & & & J1037-2934 & & J1037-2934 \\ 
\enddata
\tablecomments{ $(a)$ Bandpass calibrators ; $(b)$ Phase calibrators ; $(c)$ Amplitude calibrators \\
PI names: (1) Kelsey Johnson ; (2) Amy Reines ; (3) Nia Imara ; (4) Sara Beck ; (5) Zhiyu Zhang}
\end{deluxetable*}

\vspace{-.5cm}
\subsection{Spectral Line Image Cubes \label{sec:imaging_spec}}

Spectral line image cubes were generated using CASA 6.5.1 with the continuum-subtracted spectral line visibilities. The continuum subtraction was performed using the CASA task \textit{uvcontsub}, which uses line-free channels and a linear fit of order 1 to estimate the continuum. The continuum-subtracted visibilities were then imaged using the CASA task \textit{tclean} with the hogbom algorithm for deconvolution of the point spread function (PSF), \textit{briggsbwtaper}, and \textit{perchanweightdensity}. These options provide a flat root mean square (RMS) noise and beam size distribution over the channels, and the beam size comparable to that of the continuum image. The velocity frame is set to BARY, which normalizes the velocity to the barycentric system. The cleaning process was repeated until the maximum of the residual map reached 2$\sigma$, where $\sigma$ is the RMS noise of the image cube per channel. The line-free channels for the continuum subtraction, image size, cell size, velocity spacing, robust value for the \textit{briggsbwtaper}, and rest frequency are summarized in Table~\ref{tab:stat_images}. 
The resulting image cubes were smoothed to the largest beam size within the selected velocity range including lines with the CASA task \textit{imsmooth}. The resulting beam sizes and RMS noises of molecular line image cubes are summarized in Table~\ref{tab:stat_images}.

\begin{sidewaystable}
\tablenum{2}
\centering
\begin{threeparttable}
\caption{Parameters for Imaging and Image Statistics}
\label{tab:stat_images}
\begin{tabular}{ccccccccc}
\hline
\colhead{(1)} & \colhead{(2)} & \colhead{(3)} & \colhead{(4)} & \colhead{(5)} & \colhead{(6)} & \colhead{(7)} & \colhead{(8)} & \colhead{(9)} \\ 
\colhead{\multirow{2}{*}{Line}} & \colhead{Line-free} & \colhead{\multirow{2}{*}{Image size}} & \colhead{\multirow{2}{*}{Cell size}} & \colhead{\multirow{2}{*}{Velocity spacing}} & \colhead{\multirow{2}{*}{robust}} & \colhead{\multirow{2}{*}{Rest frequency}} & \colhead{Synthesized} & \colhead{\multirow{2}{*}{RMS}}\\
\colhead{} & \colhead{Channels} & \colhead{} & \colhead{} & \colhead{} & \colhead{} & \colhead{} & \colhead{beam size} & \colhead{} \\
\colhead{} & \colhead{}  & \colhead{(pixel$\times$pixel)} &\colhead{(\arcsec\ pixel$^{-1}$)}  &\colhead{(km s$^{-1}$)}  &\colhead{}  &\colhead{(GHz)} & \colhead{} & \colhead{($\mu$Jy beam$^{-1}$)}\\ 
\hline
\multirow{2}{*}{HCN (1--0)} & 301--750 & \multirow{2}{*}{512$\times$512} & \multirow{2}{*}{0.25} & \multirow{2}{*}{10.0} & \multirow{2}{*}{0.5} & \multirow{2}{*}{88.6316} & \multirow{2}{*}{1\farcs78$\times$1\farcs68} & \multirow{2}{*} {350$^{a}$} \\
& 1150--3539 & & & & & & & \\
\hline
\multirow{2}{*}{HCO$^{+}$ (1--0)} & 300--3300 & \multirow{2}{*}{512$\times$512} & \multirow{2}{*}{0.25} & \multirow{2}{*}{10.0} & \multirow{2}{*}{0.5} & \multirow{2}{*}{89.1885} & \multirow{2}{*}{1\farcs84$\times$1\farcs68} & \multirow{2}{*}{350$^{a}$} \\
& 3500--3539 & & & & & & & \\
\hline
\multirow{2}{*}{HCN (3--2)} & 200--1600 & \multirow{2}{*}{1024$\times$1024} & \multirow{2}{*}{0.025} & \multirow{2}{*}{20.0} & \multirow{2}{*}{1.5} & \multirow{2}{*}{265.886} & \multirow{2}{*} {0\farcs19$\times$0\farcs16} & \multirow{2}{*} { $<$ 470$^{a}$}\\
& 2000--3640 & & & & & & & \\
\hline
\multirow{2}{*}{HCO$^{+}$ (3--2)} & 200--1600 & \multirow{2}{*}{1024$\times$1024} & \multirow{2}{*}{0.025} & \multirow{2}{*}{7.5} & \multirow{2}{*}{1.5} & \multirow{2}{*}{267.558} & \multirow{2}{*}{0\farcs18$\times$0\farcs16} & \multirow{2}{*}{325$^{a}$}\\
& 2500--3640 & & & & & & & \\
\hline
\multirow{2}{*}{CO (1--0)} & 100--1700 & \multirow{2}{*}{1024$\times$1024} & \multirow{2}{*}{0.05} & \multirow{2}{*}{1.693} & \multirow{2}{*}{0.0} & \multirow{2}{*}{115.208201} & \multirow{2}{*}{0\farcs45$\times$0\farcs28} & \multirow{2}{*}{1030$^{a}$}\\
& 2150--3740 & & & & & & & \\
\hline
\multirow{2}{*}{CO (3--2)} & 10--90 & \multirow{2}{*}{1024$\times$1024} & \multirow{2}{*}{0.05} & \multirow{2}{*}{1.693} & \multirow{2}{*}{0.5} & \multirow{2}{*}{345.79599} & \multirow{2}{*}{0\farcs31$\times$0\farcs28} & \multirow{2}{*}{1000$^{a}$}  \\
& 161--230 & & & & & & & \\
\hline
99~GHz & -- & 1024$\times$1024 & 0.05 & -- & 0.0 & -- & 0\farcs55$\times$0\farcs32 & 13.0 \\
\hline
113~GHz & -- & 1024$\times$1024 & 0.05 & -- & 0.0 & -- & 0\farcs45$\times$0\farcs27 & 18.0 \\
\hline
251~GHz & -- & 1024$\times$1024 & 0.03 & -- & 1.5 & -- & 0\farcs18$\times$0\farcs15 & 12.4 \\
\hline
340~GHz & -- & 1024$\times$1024 & 0.04 & -- & 0.5 & -- & 0\farcs32$\times$0\farcs27 & 27.7 \\
\hline
\end{tabular}
\begin{tablenotes}
\item $^{a}$ The unit is ($\mu$Jy beam$^{-1}$ (channel width)$^{-1}$) for spectral image cube.
\end{tablenotes}
\end{threeparttable}
\end{sidewaystable}

\subsection{Continuum Imaging}

We created four different continuum images: low-frequency Band 3 (low-Band 3), high-frequency Band 3 (high-Band 3), Band 6, and Band 7 continuum images. The Band 3 continuum data are from ALMA project codes 2011.0.00348.S, 2015.1.01569.S, 2016.1.00027.S, and 2019.1.01641.S, which were split into low- and high-frequency data at 110~GHz. This frequency was chosen to maximize the angular resolution of the high-frequency Band 3 continuum image. The Band 6 and 7 continuum data are from ALMA project codes 2012.1.00413.S and 2016.1.00492.S, respectively.

The continuum images were created by combining the visibilities from the continuum spectral windows (SPWs) with wide channel widths of 15.625~MHz and the line-free channels from the SPWs containing spectral lines with narrow channel widths (as explained in Section~\ref{sec:imaging_spec}). 
The latter were smoothed to a channel width of 15.625~MHz in SPWs for the continuum using the CASA task \textit{mstransform}. The continuum images were generated using the CASA 6.5.1 task \textit{tclean} with the briggs weighting function and the hogbom deconvolution algorithm. The standard multi-frequency synthesis (MFS) was applied for imaging the continuum data, except for the low-Band 3. For the low-Band 3 data, we applied the multi-term MFS \citep{Rau2011}, as its fractional bandwidth is about 23.7\% (with a bandwidth of 23.4~GHz, or 87.1--110.5 GHz, relative to the effective frequency of 98.8~GHz). We used \textit{nterms}=3, which yields a more accurate intensity map by calculating the mean intensity, spectral index, and curvature over the frequency. The cleaning was stopped when it reached 2$\sigma$, where $\sigma$ is the RMS noise of the image. We summarized the parameters used in the CASA task \textit{tclean} and the resulting synthesized beam sizes and RMS noise in Table~\ref{tab:stat_images}.

The synthesized beam sizes and sensitivities of the continuum images in this paper are slightly different from those obtained in \citet{Costa2021} for 113 and 340~GHz, while the characteristics of the 251~GHz image are the same. Our 113~GHz continuum image was obtained by concatenating the data of 2015.1.01569.S, 2016.1.00027.S, and 2019.1.01641.S, which is different from \citet{Costa2021} who used 2016.1.00027.S for the 113 GHz continuum image. The synthesized beam size and RMS noise were 0\farcs38$\times$0\farcs21 and 17~$\mu$Jy beam$^{-1}$, respectively \citep{Costa2021}, where the beam size is 18\% better than our 113~GHz continuum image. This difference is attributed to adding two other observations to our data, 2015.1.01569.S and 2019.1.01641.S, since the robust value used in CASA task \textit{tclean} was the same as ours. The project codes 2015.1.01569.S and 2019.1.01641.S have maximum baselines of 1.4~km and 181.9~m, respectively, while the project 2016.1.00027.S has a maximum baseline of 2.6~km. Adding these two data with shorter baselines degraded the angular resolution when the data of 2016.1.00027.S is used solely.
While the synthesized beam sizes are almost same for the 340 GHz images, the RMS noises are 27.7 and 52.0 $\mu$Jy beam$^{-1}$ for our image and that of \citet{Costa2021}. The reason for this difference is unclear.

\subsection{Astrometry}

It is important to facilitate the comparison of multi-wavelength observations of Henize 2--10 taken with different telescopes. The location of the BH in He 2--10 was originally determined in the reference frame of 2MASS, due to its absolute positional accuracy of $\lesssim$ 0\farcs1. Images from {\it HST}, the VLA, and {\it Chandra} were shifted to match the locations with the astrometric frame of {it 2MASS} \citep{Reines2011}. Here, we identified the offsets in our ALMA observations with respect to the 2MASS reference frame and applied them to the ALMA observations.  

We used the highest angular resolution ALMA continuum image at 251~GHz (with a synthesized beam size of 0\farcs18$\times$0\farcs16) and the 1.4~GHz continuum image observed at the Long Baseline Array (LBA) with a synthesized beam size of 0\farcs11$\times$0\farcs03 to estimate the offset. The astrometry of the LBA 1.4~GHz continuum image was corrected to that of the 2MASS \citep{Reines2012}. The shift was determined by matching the local maximum in the 251~GHz continuum image to the center of the BH in the 1.4~GHz continuum image. We estimated the shift of $\rm \Delta RA=$0\farcs0735 West and $\rm \Delta DEC=$0\farcs095 North and applied this shift to all ALMA spectral line image cubes and continuum images.

\section{Analysis and Results \label{sec:results}}

\subsection{Continuum Emission and Radio SED of the Black Hole}

\begin{figure}
    \centering
    \includegraphics[width=4.0in]{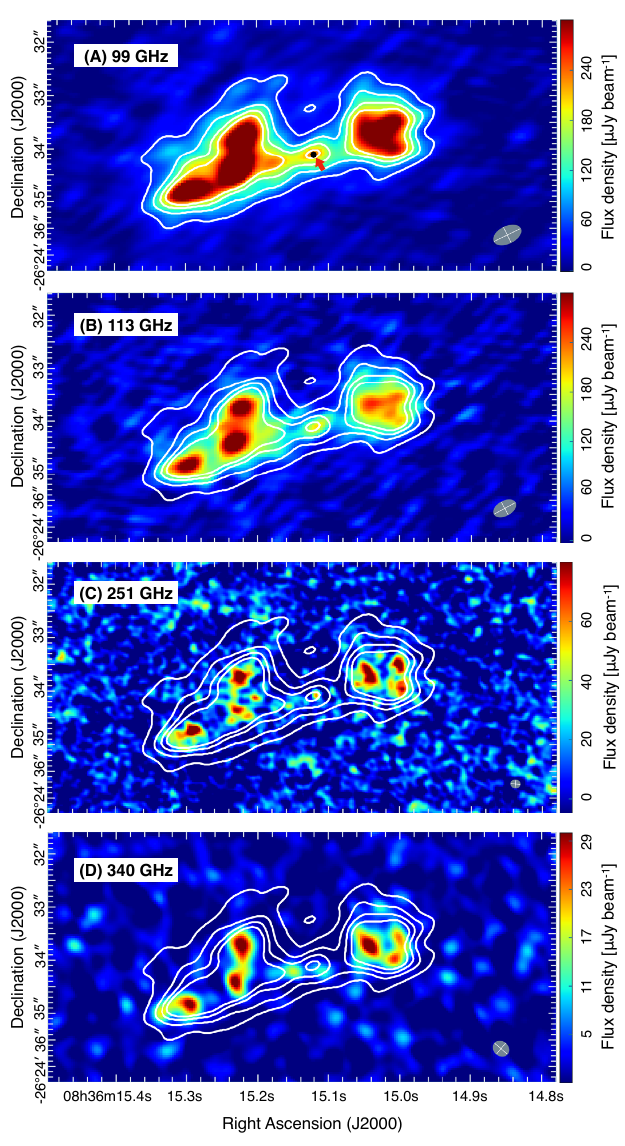}
    \caption{Continuum images at 99~GHz (A), 113~GHz (B), 251~GHz (C), and 340~GHz (D). The central BH is marked with the black ellipse and red arrow in the 99~GHz continuum image (panel A). White contours represent the 99~GHz continuum at levels of 50, 100, 150, and 200 $\mu$Jy beam$^{-1}$. The beam size in each image is shown as the gray ellipse at the bottom right. }
    \label{fig:continuum}
\end{figure}

Figure~\ref{fig:continuum} presents the sub-mm continuum maps at 99 (first row), 113 (second row), 251 (third row), and 340~GHz (fourth row), respectively. The white contours show the 99~GHz continuum at 100, 150, 180, 200, and 250 $\mu$Jy beam$^{-1}$. The local maxima of the contours in the center indicate the position of BH, which coincides with the positions determined by the VLA \& LBA \citep{Reines2011, Reines2012}. The continuum images show two prominent blobs and one central source. The two strong emission blobs correspond to the starburst regions, and the central source indicates the massive BH \citep{Reines2011, Reines2012}. The position of the BH is defined in the continuum images as the local maxima in the central region between the two prominent starburst regions, as indicated in the top panel of Figure~\ref{fig:continuum}. The two starburst regions are clearly visible in all the continuum maps; however, the central BH is not seen in the 340~GHz continuum map. In the 340~GHz continuum image, the BH is located between two small continuum sources located 0\farcs217 (10~pc) and 0\farcs526 (23~pc) away.

An earlier study by \citet{Allen1976} showed that He~2--10 as a whole is dominated by nonthermal emission. However, high angular resolution VLA and ALMA observations have shown that the central few hundred parsecs are dominated by free-free emission from bursty star formation \citep{Johnson2003, Johnson2018, Costa2021}. The central BH, on the other hand, is governed by nonthermal synchrotron emission \citep{Johnson2003, Reines2011, Reines2012, Costa2021}. Although the VLA + ALMA SED analysis for the BH has been done by \citet{Costa2021}, we revisited this analysis with revised flux density measurements of the submillimeter (sub-mm) continuum and inclusion of 1.4~GHz flux density observed at the LBA \citep{Reines2012}. 

The radio SED of the BH from 1.4 to 340 GHz is shown in Figure \ref{fig:BH_sed}. In addition to our new ALMA measurements in the sub-mm regime (see below), we include measurements from the literature including 1.4~GHz data from the LBA \citep{Reines2012}, 5 and 8~GHz data from the VLA \citep{Reines2011}, as well as 15, 22, and 33~GHz data from the VLA \citep{Costa2021}. Here, the flux density at 33~GHz is not included in this analysis because the data suffered from decorrelation due to weather conditions and the low declination of He 2--10 at this high frequency \citep{Costa2021}. Therefore, the flux density at 33~GHz is presented with a lower limit (upper arrow) in Figure~\ref{fig:BH_sed}.

The flux densities of the BH in the sub-mm regime were estimated with our reimaged ALMA continuum data. First, to help mitigate the impact of different beam sizes, we convolved all the continuum images to have the same synthesized beam size as the lowest one (0\farcs55$\times$0\farcs32 for the 99~GHz image) using the CASA task \textit{imsmooth}. While the inherent limitation of interferometry may still result in different spatial sensitivities among the observations, as well as the omission of large-scale emission, this would not be a major concern if the emission associated with the BH is point-like. However, if the emission is marginally resolved, we may be missing flux density that could, for example, lead to the low measurement at 251 GHz seen in Figure \ref{fig:BH_sed}.

We performed aperture photometry on the sub-mm continuum using an elliptical aperture equivalent to one beam size centered on the BH position (i.e., 0\farcs55$\times$0\farcs32 with PA$=$-64.0\degree\ marked with a blue ellipse in Figure~\ref{fig:BH_sed}). The arrow points to the BH which is shown as a black ellipse. The measured flux densities of the BH are 140$\pm$14, 134$\pm$21, 29$\pm$46, and 57$\pm$48~$\mu$Jy at 99, 113, 251, and 340~GHz, respectively. Here, it is worth noting that the uncertainty of the flux density may be underestimated because we measured it as a mean of the RMS noise within an aperture. The RMS noise maps were produced with the AIPS task \textit{RMSD} \citep{Greisen2003}, which calculates the RMS of signals after rejecting signals greater than 3$\sigma$ within 150 pixels of each pixel.

\begin{figure}
    \centering
    \includegraphics[width=7.0in]{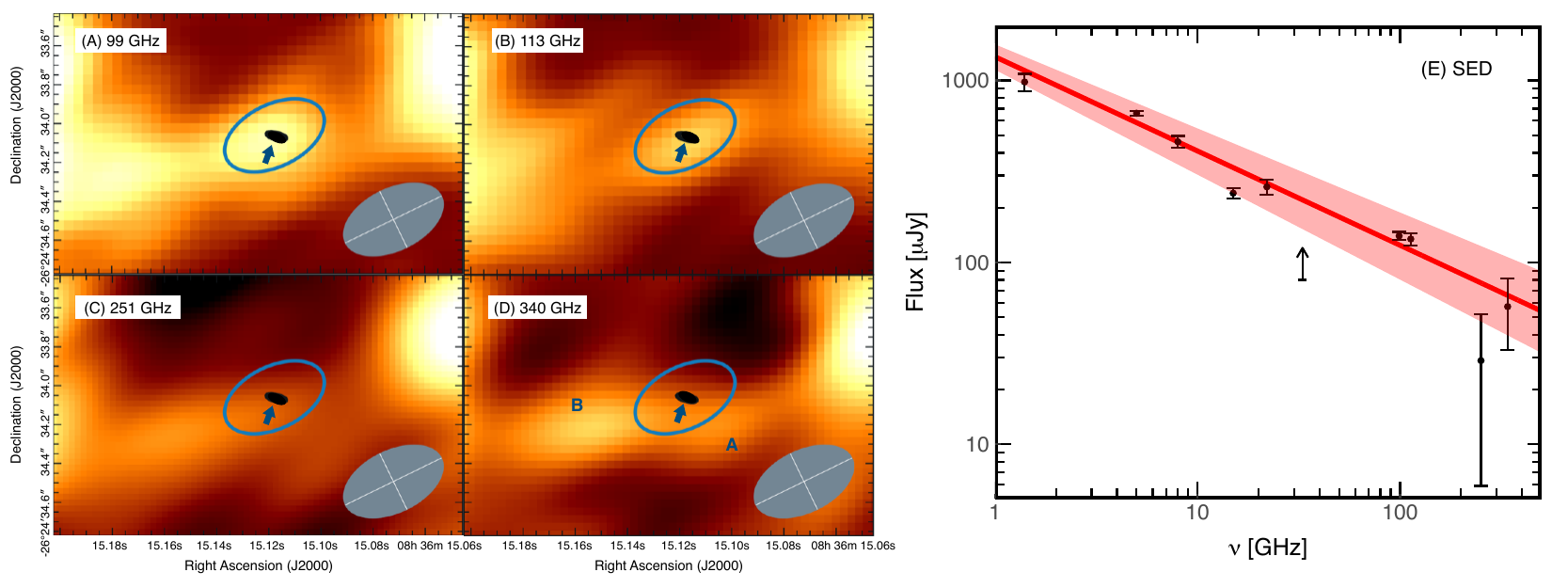}
    \caption{Continuum images at 99~GHz (A), 113~GHz (B), 251~GHz (C), 340~GHz (D), and the SED analysis (E). Continuum images were convolved to have the same synthesized beam size of 0\farcs55$\times$0\farcs32. The central BH is marked with the small black ellipse indicated by the arrow. The blue ellipse is the aperture of our flux density measurements. The beam size in each image is shown as the gray ellipse at the bottom right. The source A and B in the 340~GHz continuum image (panel D) are adjacent sources of the BH. In panel (E), the flux densities are depicted as black data points, with a black arrow indicating the lower limit at 33 GHz. The best-fit model is represented by a red line, and the 1$\sigma$ range of the best-fit model is delineated by the red shaded region, providing a measure of the model's uncertainty.}
    \label{fig:BH_sed}
\end{figure}

Our measurements are consistent with those of \citet{Costa2021} at 113~GHz, but not at 251 and 340~GHz. \citet{Costa2021} reported flux densities of 70$\pm$30, 130$\pm$30 $\mu$Jy at 258 and 340~GHz, respectively, while our measurements are 29$\pm$46 and 57$\pm$48 $\mu$Jy at these frequencies. At 251~GHz, our photometry shows 2.4 times less flux density than that of \citet{Costa2021}. \citet{Costa2021} likely measured the flux density of a point source shown within the contour in the center of panel (C) of Figure~\ref{fig:continuum}. However, this point source is smeared after convolution to the beamsize of 0\farcs55$\times$0\farcs32 in our measurement, as shown in panel (C) of Figure~\ref{fig:BH_sed}. Since we applied aperture photometry on the convolved image, we measured a smaller flux density. At 340~GHz, our measurement is 2.3 times fainter than the flux density obtained by \citet{Costa2021}. This difference is due to the difference in measurements. There are two continuum sources near the BH position with distances of 0\farcs22 (source A) and 0\farcs53 (source B) from the BH, as shown in panel (D) of Figure~\ref{fig:BH_sed}. Their flux densities were measured as 125$\pm$80 and 342$\pm$130 $\mu$Jy for the sources A and B, respectively, using PyBDSF \citep{Mohan2015}. \citet{Costa2021} likely reported the flux density of source A as that of the BH. Our aperture for the BH does not include source A and source B entirely, but includes a part of source A, as seen in panel (D) of Figure~\ref{fig:BH_sed}. Therefore, our reported values are more robust than the ones reported by \citet{Costa2021}.

Our measurements have larger uncertainties than those of \citet{Costa2021} at both 251 and 340~GHz. These larger uncertainties are due to the convolution to the beamsize of 0\farcs55$\times$0\farcs32. The differences in beam area between the original and convolved images are 6.5 and 2.0 at 251 and 340~GHz, respectively. Convolution to the larger beam size introduces a larger uncertainty. Quantitatively, the 251~GHz image with the original synthesized beam size of 0\farcs18$\times$0\farcs14 has an RMS noise of 12.4 $\mu$Jy beam$^{-1}$, but its RMS noise increases to 24.7 $\mu$Jy beam$^{-1}$ after convolution to a synthesized beamsize of 0\farcs55$\times$0\farcs32. Similarly, the 340~GHz image has an RMS noise of 27.7 $\mu$Jy beam$^{-1}$ originally, but increases to 38.7 $\mu$Jy beam$^{-1}$ after convolution. Our measurements at 251 and 340~GHz had larger uncertainties, but it is important to match the beamsizes across the frequencies in order to accurately measure the spectral index. Additionally, it is important to measure the flux density of the BH at the correct position, rather than the flux density of a nearby yet different source. As a result, the flux densities we report are more accurately measured than previous values. 

In order to estimate the best-fit parameters and corresponding uncertainties for an SED ranging from 1.4 to 340~GHz, we employed the R function \textit{nls} for the non-linear least square fitting and utilized \textit{confint2} to estimate the uncertainties of the best-fit parameters. Four models were tested, including i) a three-component model with non-thermal, free-free, and dust emission, $S \sim S_{0, nth} \nu^{\alpha_{nth}} + S_{0, ff} \nu^{-0.1} + S_{0, dust} \nu^{4}$, ii) a two-component model with non-thermal and free-free emission, $S \sim S_{0, nth} \nu^{\alpha_{nth}} + S_{0, ff} \nu^{-0.1}$, iii) a two-component model with non-thermal and dust emission, $ S \sim S_{0,nth} \nu^{\alpha_{nth}} + S_{0, dust} \nu^{4}$, and iv) a one-component model with non-thermal emission, $S \sim S_{0,nth} \nu^{\alpha_{nth}}$. The models that included dust emission did not yield physical results. The two-component model with non-thermal and free-free emission had large uncertainties and poorly constrained solutions. The one-component model with non-thermal emission provided a meaningful solution, with the best-fit model of $S= (1349 \pm 217) \nu^{-0.52 \pm 0.06}$~$\mu$Jy, represented by the red line with a light red shaded region in panel E of Figure~\ref{fig:BH_sed}.

\subsection{Molecular Line Emission}\label{sec:mol}

\subsubsection{Moment Maps}

\begin{figure}
    \centering
    \includegraphics[width=6.8in]{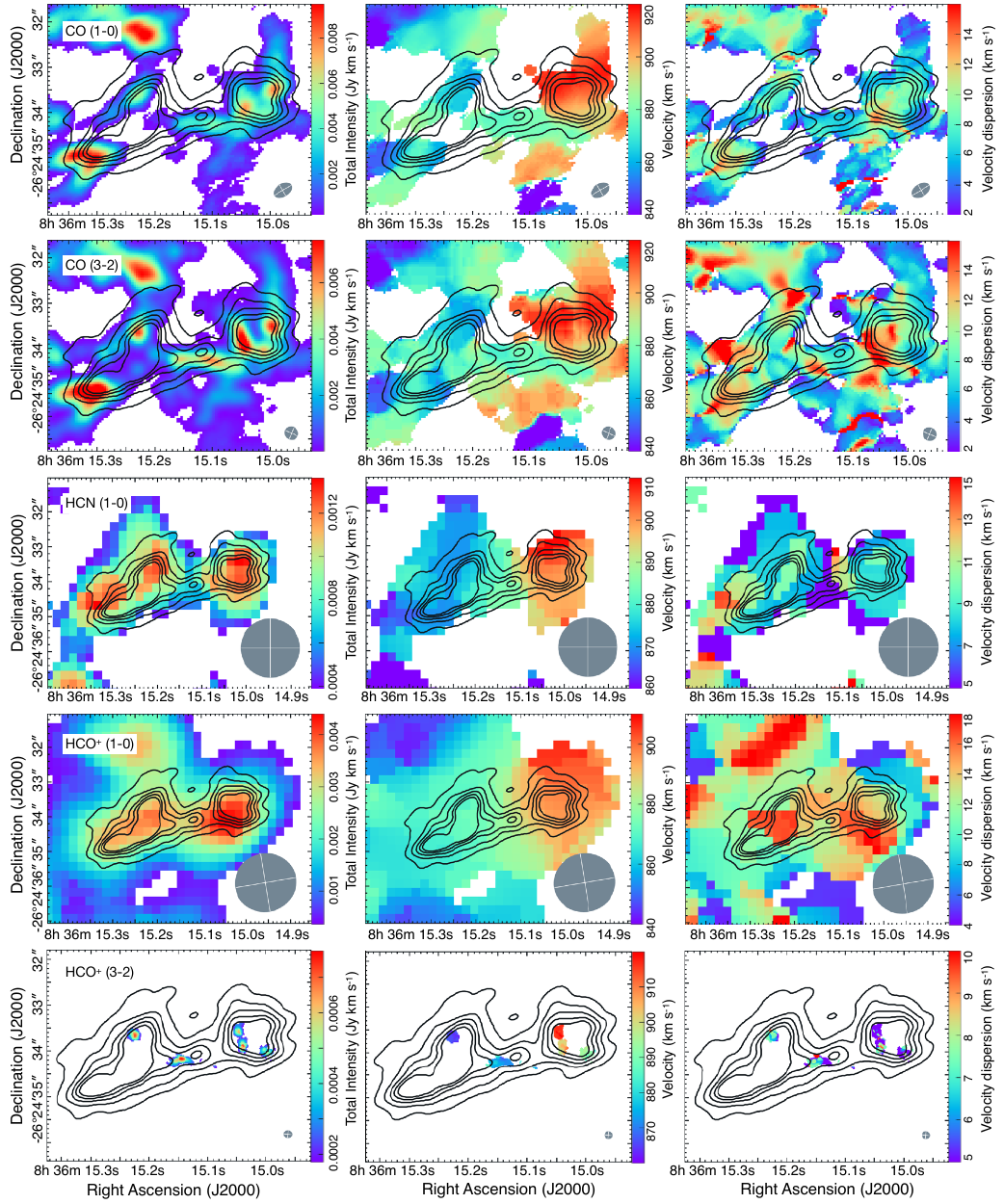}
    \caption{Moment maps of CO (1--0), CO (3--2), HCN (1--0), HCO$^{+}$ (1--0), and HCO$^{+}$ (3--2) from top to bottom. Total intensity, velocity field, and velocity dispersion maps are shown in the left, middle, and right panels, respectively. The black contours indicate the flux density of the 99~GHz continuum, corresponding to 50 (5$\sigma$), 100 (10$\sigma$), 150 (15$\sigma$), and 200 (20$\sigma$) $\mu$Jy per beam size of 0\farcs55$\times$0\farcs3. The BH is located in the center of the innermost black contour of the continuum. The beam sizes are marked with solid ellipses in the bottom right corners.}
    \label{fig:moment_maps}
\end{figure}

Moment maps of molecular gas can provide insight into the morphology and kinematics of the gas. Moment maps were generated using the 3D-Based Analysis of Rotating Objects via Line Observations \citep[3D-Barolo; ][]{Teodoro2015}, which employs the Duchamp algorithm to identify pixels above a primary signal-to-noise ratio (S/N) and agglomerates neighboring pixels above a secondary S/N. The initial moment maps were created by running 3D-Barolo on the non-primary beam (PB) attenuation corrected image cubes, and the final moment maps were derived by multiplying the PB attenuation corrected image cubes by the mask image cubes created during the generation of the initial moment maps. For this analysis, the primary and secondary S/N were set at 5 and 3, respectively.

In Figure~\ref{fig:moment_maps}, the moment maps of the CO (1--0), CO (3--2), HCN (1--0), HCO$^{+}$ (1--0), and HCO$^{+}$ (3--2) lines are shown from top to bottom. The total intensity (moment 0), velocity field (moment 1), and velocity dispersion (moment 2) maps are displayed in the left, middle, and right panels, respectively. The black contours indicate the 99~GHz continuum, with the BH located at the local maximum between two major blobs of star formation. The moment maps of the molecular gas lines indicate that there is a significant amount of molecular gas beyond the continuum regions traced by the black contours. These extended features are consistent with previous observations \citep{Baas1994, Beck2018, Imara2019}. Total intensity maps of the molecular gas (left panels of Figure~\ref{fig:moment_maps}) show two prominent regions corresponding to the strong continuum blobs of Figure~\ref{fig:continuum}. The maximum intensities of the molecular gas in the left panels of Figure~\ref{fig:moment_maps} coincide with the maximum positions in the continuum. The BH is located between the two prominent starburst regions in the molecular gas distributions, forming a spur-like structure visible in the high-resolution CO data. The detection of HCO$^{+}$ (1--0), HCO$^{+}$ (3--2), and HCN (1--0) above S/N $>$ 5 indicate the existence of significant dense gas (with critical density $\gtrsim~10^{6}$~cm$^{-3}$) in the vicinity of the BH.

\subsubsection{Position-Velocity Diagram}

\begin{figure}
    \centering
    \includegraphics[width=6.1in]{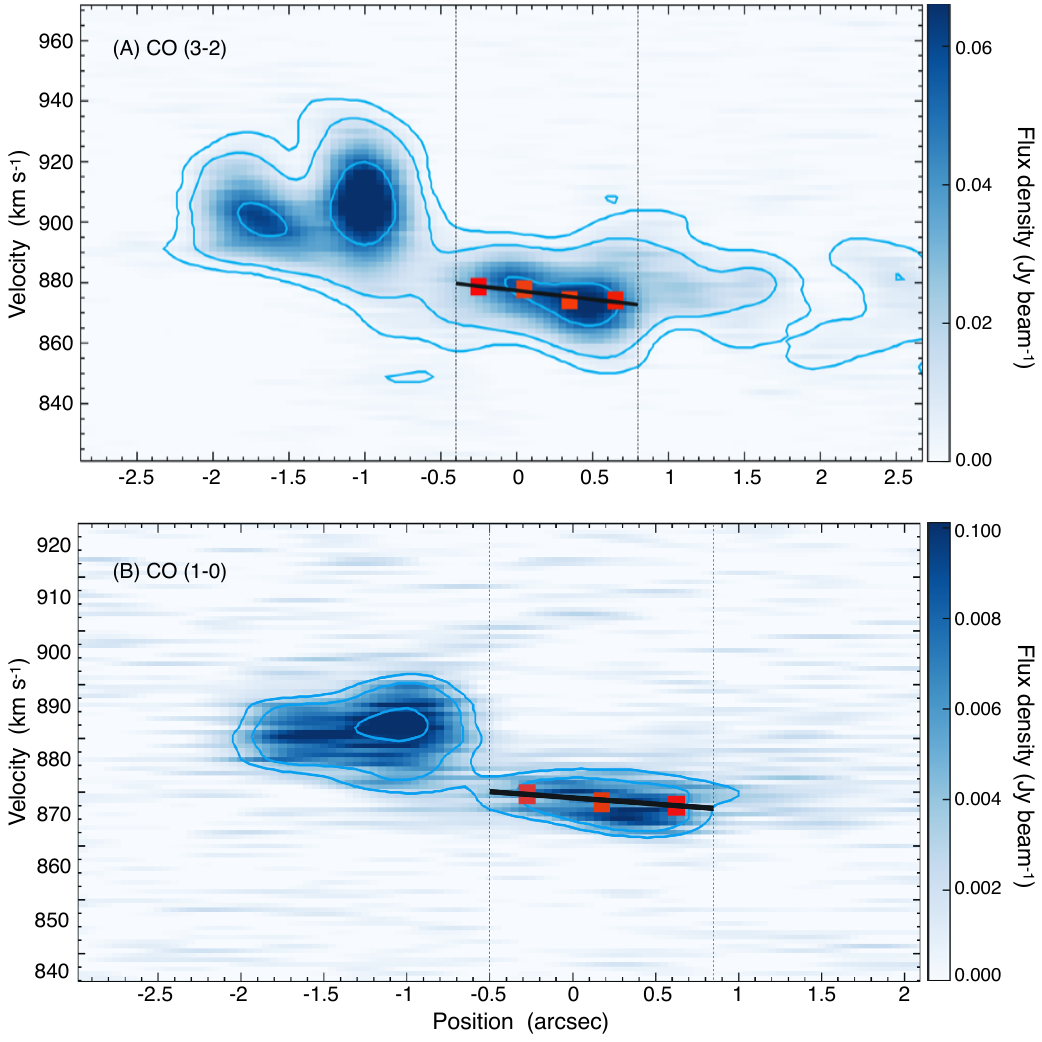}
    \caption{PVDs of CO J$=$3--2 (panel A) and CO J$=$1--0 (panel B) are presented with the weighted mean velocity in each position bin above 3$\sigma$ (red squares) and the velocity gradient (black lines). The light blue contours indicate 3, 10, and 50$\sigma$ in CO (3--2) and 3, 5, and 10$\sigma$ in CO (1--0), respectively. The dotted vertical lines present the regions estimating the velocity gradient, e.g., -0\farcs4 $< x <$ 0\farcs8 for CO (3--2) and -0\farcs5 $< x <$ 0\farcs85 for CO (1--0).}
    \label{fig:pvd}
\end{figure}

The position-velocity diagram (PVD) serves as a valuable tool for elucidating the kinematics of gas. The ionized gas in the central region of He 2--10 exhibits a sinusoidal motion in the PVD, which is well fit by the bipolar outflow model presented in \citet{Schutte2022}. \citet{Schutte2022} used \textit{HST}/STIS spectroscopy, employing a long slit centered on the BH with a width of 0\farcs2 and a position angle of 100\degree\ aligned with the ionized filament seen in H$\alpha$ imaging. In our investigation, we look for a similar pattern in the velocities of the molecular gas by generating PVDs along a line with a position angle of 100\degree, centered on the BH, from our CO (3--2) and CO (1--0) image cubes. We extracted PVDs along the line in bins corresponding to the synthesized beam sizes of CO (3--2) and CO (1--0) image cubes, i.e., 0\farcs3 and 0\farcs45, respectively. However, these values are more than 3$\times$ larger than the spatial resolution of the {\it HST}/STIS observations ($\sim 0\farcs1$).

Figure~\ref{fig:pvd} illustrates the resulting PVDs for CO (3--2) in panel (A) and CO (1--0) in panel (B). Unlike the ionized gas observed by \citet{Schutte2022}, no sinusoidal motions were discerned in either PVD. This may be attributed to the coarser angular resolution of our observations, leading to a beam smearing effect that obscures outflow signals amid adjacent background emissions. Alternatively, the absence of sinusoidal motion in the molecular gas could signify a genuine lack of such motion, suggesting a potential insensitivity of the molecular gas to the BH outflow, particularly if the BH outflow lacks sufficient strength \citep{Koudmani2022}. 

We calculated the velocity gradient from the PVD of the CO (3--2) line in panel (A) of Figure~\ref{fig:pvd}, leveraging the smaller beam size of the CO (3--2) image cube compared to that of CO (1--0). We restrict this analysis to the gas in the central region near the BH between -0\farcs4 ($-18$~pc) to 0\farcs8 ($35$~pc) with velocities in the range $\sim$870--900 \kms. The resulting weighted mean velocity (red squares in Figure~\ref{fig:pvd}), determined above 3$\sigma$ with a bin width of 0\farcs3 (where $\sigma=1.0$~mJy beam$^{-1}$ (1.693 \kms)$^{-1}$, yielded a velocity gradient of $-5.8$~km s$^{-1}$ \arcsec$^{-1}$. We repeated this analysis for CO (1--0), constructing a PVD along a line with a position angle of 100\degree\ and a width of 0\farcs45 (matching the beam size of CO (1--0)), resulting in a similar velocity gradient of $-4.6$ km s$^{-1}$ \arcsec$^{-1}$).

Our derived velocity gradient is shallower than the $-7.8$~km s$^{-1}$ \arcsec$^{-1}$ reported by \citet{Imara2019}. The discrepancy can be attributed to differences in i) the length of the position axis used to estimate the velocity gradient, ii) the position angle, and iii) the bin size used to trace gas velocity. \citet{Imara2019} estimated the velocity gradient over a range of -1\farcs7 ($\sim$70~pc) to 1\farcs7 along a line with a position angle of 143\degree\ and a width of 4\arcsec. In our PVD (Figure~\ref{fig:pvd}), molecular gas at approximately 1\farcs5 is associated with the western star-forming region, contributing to a steeper velocity gradient due to its high velocity. The differences in position angle and bin size introduce additional inconsistencies, particularly given the asymmetrical distribution of molecular gas around the BH.

\subsubsection{Molecular Gas in the Vicinity of the BH}

\begin{figure}[!h]
    \centering
    \includegraphics[width=7.0in]{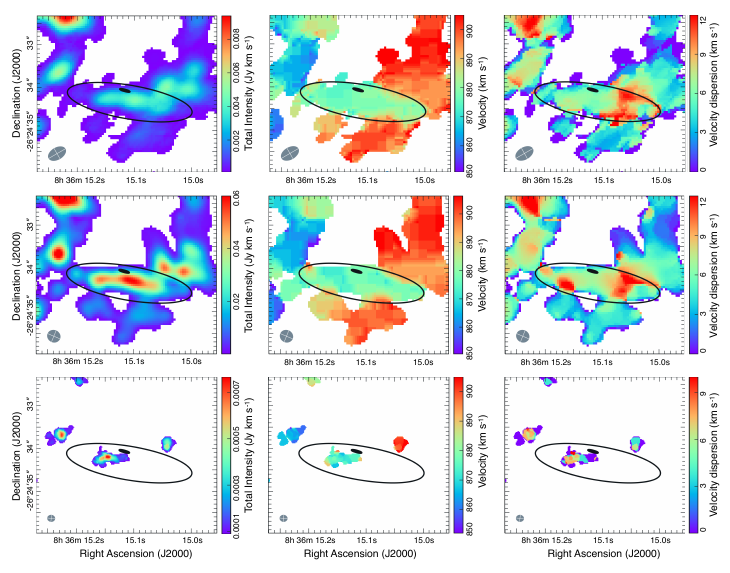}
    \caption{Moment maps of the molecular gas species observed at high angular resolution including CO (1--0), CO (3--2), and HCO$^{+}$ (3--2) (top to bottom), where the moment maps are generated with a restrictive velocity range of 852.1--904.6~km s$^{-1}$ (see \S \ref{sec:mol}). The total intensity, velocity field, and velocity dispersion are presented in the left, middle, and right panels, respectively. The small black contour indicates the position of BH traced by the high brightness temperature radio core detected with the LBA \citep{Reines2012}. The outer black ellipse defines the region we refer to as the BH vicinity, which has a near uniform velocity that is consistent with the systemic velocity of the galaxy, $V_{sys}=873$~km s$^{-1}$ \citep{Kobulnicky1995}.}
    \label{fig:BH_mom}
\end{figure}

In this work, we are primarily interested in examining the molecular gas in the vicinity of the BH\footnote{The ALMA observations do not have sufficient angular resolution to probe the BH sphere of influence. Given the BH mass of $\rm \sim 10^{6} \; M_{\odot}$ \citep{Reines2016, Riffel2020}, the sphere of influence is $\sim 4$~pc (0\farcs1), which is at least three times smaller than the angular resolution of our observations.}. We define this region (i.e., the BH vicinity) as follows. There is a distinct velocity gas blob near the BH in HCO$^{+}$ (3--2), seen in the bottom middle panel of Figure~\ref{fig:moment_maps}.
The velocity range of this blob is 852.1--904.6~km s$^{-1}$ in HCO$^{+}$(3--2). To help isolate the gas in the vicinity of the BH, both spatially and in velocity, we generated moment maps of the molecular gas tracers with higher angular resolution (CO (1--0), CO (3--2), and HCO$^{+}$ (3--2)) within this restricted velocity range (Figure \ref{fig:BH_mom}). 

The resulting moment maps reveal an elongated CO gas structure with a distinct velocity from the surrounding regions, as seen in the middle panels of Figure~\ref{fig:BH_mom}. This distinct velocity region is defined as the BH vicinity, represented by the black ellipse in Figure 6. The ellipse is centered at RA 08$^{h}$ 36$^{m}$ 15.108$^{s}$ and Dec $-26$\degree\ 24\arcmin\ 34.343\arcsec, with a size of 2\farcs97$\times$0\farcs79 ($\approx$ 128~pc$\times$34~pc) and a position angle of 80.34\degree\ as presented in Table~\ref{tab:BH_vicinity}. The moment maps in Figure~\ref{fig:BH_mom} clearly demonstrate that the BH (small black contour) is located north of the BH vicinity (black ellipse) in molecular gases such as CO and HCO$^{+}$ (3--2). In other words, the BH is not at the center of the molecular gas structure. The lack of the molecular gas to the north of the BH is likely a result of star formation feedback occurring in the young massive star clusters with ages of $\sim 4-5$~Myrs \citep{Beck2018}, as seen in the H${\alpha}$ image in Figure~\ref{fig:co_ratio}. This could help explain why the BH is accreting at such a low Eddington ratio ($\sim 10^{-6}$; \citealt{Reines2016}). Furthermore, the asymmetrical distribution of molecular gas around the massive BH resembles the central molecular zone of the Milky Way \citep{Bally1988} and the central region of NGC~253 \citep{Meier2015, Krieger2019, Levy2022}. In the cases of the Milky Way and NGC~253, the origin of this asymmetry is attributed to feedback from star formation in their centers \citep{Henshaw2022, Krieger2019}.

We have measured the velocity ranges within the BH vicinity using three different molecular lines: CO J=1--0, CO J=3--2, and HCO$^{+}$ J=3--2. The velocity ranges were found to be 865--899~km s$^{-1}$, 871--904 km s$^{-1}$, and 871--904 km s$^{-1}$, respectively. Excluding the borders, the velocity ranges for CO J=1--0 and J=3--2 were reduced to 871--886 km s$^{-1}$ and 871--885 km s$^{-1}$, respectively. The mean velocities within the BH vicinity are 880$\pm$5 (CO J=1--0), 880$\pm$6 (CO J=3--2), and 875$\pm$6~km s$^{-1}$ (HCO$^{+}$ J=3--2), and are therefore consistent with one another within the uncertainties. For reference, the systemic velocity of the galaxy is 873 km s$^{-1}$ \citep{Kobulnicky1995}.
The velocity field maps do not show any significant rotation in the region defined as the BH vicinity. The velocity dispersion map of CO J=3--2 shows some high velocity dispersion along the SE and NW directions with $\Delta \sigma \approx 7$~km s$^{-1}$. The high velocity dispersion ($\Delta \sigma \approx 50$~km s$^{-1}$) of stellar motions near the BH observed by \citet{Riffel2020} is not seen in the molecular gas.

\begin{table}[!h]
    \tablenum{3}
    \centering
    \caption{Properties of molecular gas in the vicinity of the BH (black ellipse in Figure \ref{fig:BH_mom})}
    \begin{tabular}{c|c}
    \hline
        Parameter & Value \\
    \hline
        Center coordinate & RA 08$^{h}$ 36$^{m}$ 15.108$^{s}$, Dec $-26$\degree\ 24\arcmin\ 34.343\arcsec\ \\
    \hline
        Major \& minor axes & 2\farcs97$\times$0\farcs79 ($\approx$128 pc$\times$34 pc) \\ 
    \hline
        Position angle & 83.34\degree\ \\
    \hline
        \multirow{3}{*}{Mean velocity} & 880.2 $\pm$ 4.7 km s$^{-1}$ in CO (1--0)\\
                                       & 880.3 $\pm$ 5.9 km s$^{-1}$ in CO (3--2) \\
                                       & 875.3 $\pm$ 6.2 km s$^{-1}$ in HCO$^{+}$ (3--2) \\
    \hline
        H$_{2}$ mass & (0.21--1.53)$\times 10^{6}$~M$_{\odot}$\\
    \hline
    \end{tabular}
    \label{tab:BH_vicinity}
\end{table}

The molecular gas mass within the BH vicinity is estimated using the scaling relation between molecular gas mass and CO luminosity, $\rm M_{H_{2}} = \alpha_{CO} L^{'}_{CO}$. The conversion factor $\alpha_{CO}$ depends on the metallicity \citep[][ and references therein]{Bolatto2013}. \citet{Cresci2017} conducted spectroscopic observations with \textit{MUSE} to determine the metallicity of He 2--10, with a spatial sampling size of 0\farcs2$\times$0\farcs2. They created a map of metallicity distribution, which revealed significant variation across the galaxy, largely due to change in ionization parameters. Through simultaneous line-fitting on multiple lines, they found that the metallicity in the central region was super-solar, with $\rm 12+[O/H] \approx 9.0$, corresponding to $\rm Z=0.044$. Based on the relationship between $\alpha_{\rm CO}$ and metallicity outlined in \citet{Bolatto2013}, we choose a range of $\alpha_{\rm CO}$ values between 0.6 and 4.3 $M_{\odot}$ (K km s$^{-1}$ pc$^{2}$)$^{-1}$, where $\rm \alpha_{\rm CO}=4.3$ is the value for the Milky Way. Our range of $\alpha_{\rm CO}$ also includes the widely accepted value for the centers of starburst galaxies with solar metallicities, $\alpha_{\rm CO} \approx 1.1$ M$_{\odot}$ (K km s$^{-1}$ pc$^{2}$)$^{-1}$ \citep{Leroy2015, Krieger2019}. These considerations ensure that our analysis is consistent with established standards in the field. Our estimated CO luminosity is $\rm L^{'}_{CO}= 3.56 \times 10^{5} \; K\; km\; s^{-1} \; pc^{2}$, which when combined with the $\rm \alpha_{\rm CO}$ range, results in an estimated molecular gas mass of $\rm M_{H_{2}} = (0.21-1.53) \times 10^{6}$~M$_{\odot}$.

\subsubsection{CO Line Ratios}

\begin{figure}
    \centering
    \includegraphics[width=6.8in]{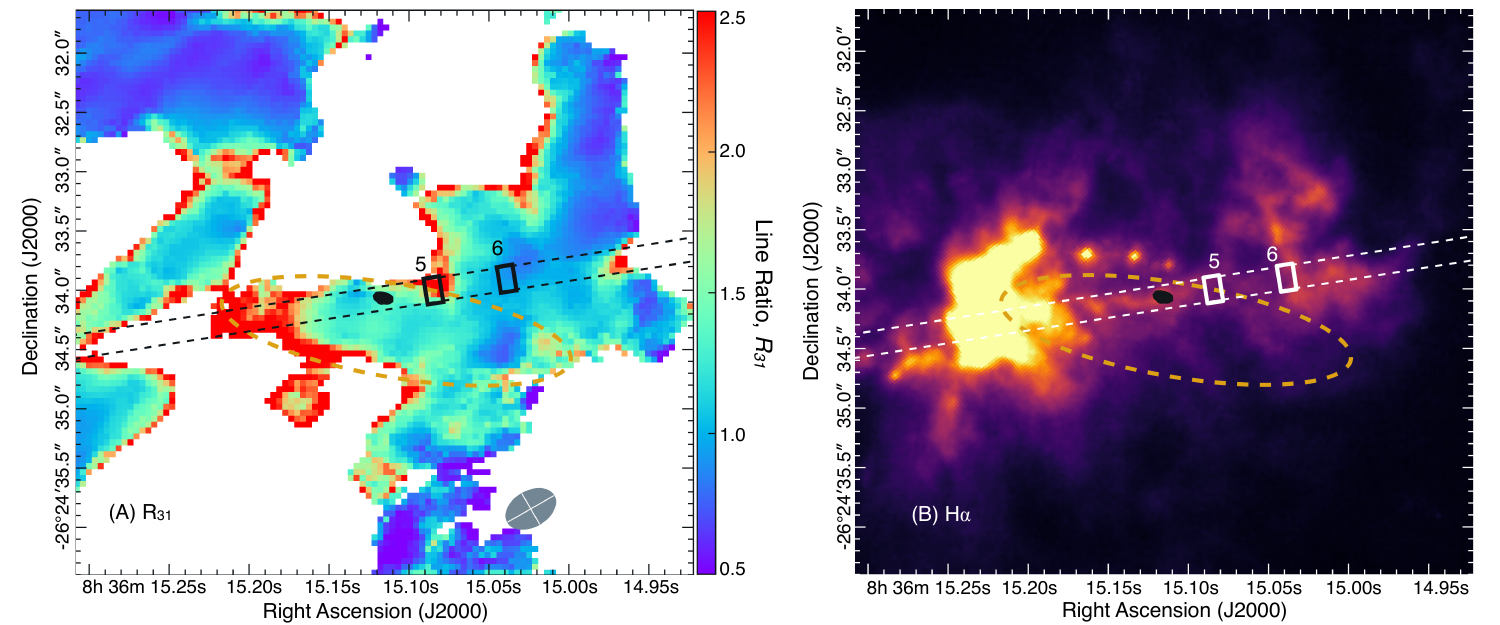}
    \caption{Left: Pixel-by-pixel line ratio map between CO (3--2) and CO (1--0), $R_{31}$. Right: {\it HST} H${\alpha}$ plus continuum image with the same field of view. The BH position is marked with black contours from the LBA 1.4 GHz observations by \citet{Reines2012}, and the region defined as the BH vicinity is indicated with the orange dashed ellipse. Regions 5 and 6 from \citet{Schutte2022} are presented as rectangles with the \textit{HST} slit position for the ionized gas outflow \citep{Schutte2022}. The highest line ratios are observed at the interface between the BH outflow and the eastern star forming region, as well as region 5 to the west of the BH. This implies the cold molecular gas in these regions is shocked by the bipolar outflow, supporting the case for positive BH feedback.}
    \label{fig:co_ratio}
\end{figure}

Line ratios between different molecular gas tracers are good indicators of gas excitation \citep{Privon2015}. The HCN (1--0) and HCO$^{+}$ (1--0) line image cubes have large beam sizes, and the HCO$^{+}$ (3--2) line image cube has a relatively low signal-to-noise ratio, so we restricted our analysis to the line ratio between CO J=1--0 and J=3--2, which have higher angular resolutions and better sensitivities. Given that the CO (3--2) line emission exhibits a more extended distribution compared to the CO (1--0) line emission, the distribution of the CO line ratio is constrained by the CO (1--0) line distribution. 
The line ratio is defined as $\rm R_{31} = L_{CO(3-2)} / L_{CO (1-0)} = (\nu_{32}/\nu_{10})^{-2} (I_{CO(3-2)}/I_{CO(1-0)})$, where $\rm I_{CO}$ is the velocity integrated line intensity in units of Jy km s$^{-1}$. Both image cubes were convolved to have the same synthesized beam size of 0\farcs45$\times$0\farcs30 using the CASA task \textit{imsmooth}. We present the pixel-by-pixel CO line ratio map in the left panel and an {\it HST} H$\alpha$ image \citep[e.g.,][]{Schutte2022} in the right panel of Figure~\ref{fig:co_ratio}. Our line ratio map of the central region of the galaxy indicates highly excited CO molecular gas as evidenced by $\rm R_{31} \gtrsim 1$, exceeding the typical range observed in the Milky Way, $\rm 0.3 < R_{31} < 0.5$ \citep{Leroy2009, Carilli2013}. This result aligns with the findings of \citet{Beck2018}, who reported a high flux ratio of $\rm S_{CO(3-2)}/S_{CO(1-0)}=6.1$ for this galaxy.

\subsubsection{Shocked Molecular Clouds from the BH Outflow}

The discovery of a low-velocity bipolar outflow originating from the BH was reported in ionized gas \citep{Schutte2022}. The precession angle and frequency of this outflow were estimated to be in the range of $2.4-6.1$\degree\ and $3.0-7.5$ revolutions per Myr, respectively. Notably, \citet{Schutte2022} identified several impacts of the BH outflow, including the stimulation of star formation in the eastern star-forming region, as well as in the dark cloud (region 5) and the dense clouds within the western star-forming region (region 6). The left panel of Figure~\ref{fig:co_ratio} reveals that the highest $\rm R_{31}$ line ratios are observed at the interface between the known ionized gas outflow from the BH and the eastern star forming region \citep{Schutte2022}, as depicted in the H$\alpha$ image (right panel of Figure~\ref{fig:co_ratio}). The line ratio at this interface is $\rm R_{31}=2.7\pm0.2$, while the region defined as the BH vicinity excluding this interface has $R_{31}=1.3\pm0.1$. Additionally, we detect a high line ratio ($\rm R_{31}=2.5\pm 0.7$) to the west of the BH at the boundary of a dark cloud of gas and dust (region 5 in Figure \ref{fig:co_ratio}). This region was previously identified by \citet{Schutte2022} and exhibits double-peaked optical emission lines, likely due to the other side of the bipolar BH outflow intercepting dense clouds and pushing them laterally (as is the case with the eastern interface region). 
The CO line ratio at region 5 is comparable to that of interface between the outflow and the eastern star forming region, consistent with this scenario. These findings suggest that the molecular gas impacted by the BH outflow to the east and west have both high density and high kinetic temperature \citep{Carilli2013}. 
It is noteworthy that the elevated $\rm R_{31}$ may be an edge effect limited by low flux densities of CO (1--0). Nevertheless, certain regions at the peripheries, particularly in the northeast and south, exhibit low $\rm R_{31}$ values. The presence of these regions implies that the the pixel-by-pixel $\rm R_{31}$ values are influenced by the relative distributions of CO (1--0) and CO (3--2) due to some physical origins, rather than being solely dictated by low CO (1--0) flux densities. Moreover, our derived $\rm R_{31}$ values are robust, exhibiting similar sensitivities, e.g., $\sigma=$1.03 mJy beam$^{-1}$ (1.693 km s$^{-1}$)$^{-1}$ for CO (1--0) and 1.00 mJy beam$^{-1}$ (1.693 km s$^{-1}$)$^{-1}$ for CO (3--2), respectively. Therefore, the elevated $\rm R_{31}$ values observed in this study are not solely constrained by low CO (1--0) flux densities.

Multiple studies have investigated the occurrence of elevated CO line ratios in different galaxies. \citet{Garcia-Burillo2014} and \cite{Viti2014} examined the CO line ratio in NGC~1068 and demonstrated that the ratio between J=3--2 and J=1--0 vary significantly between the circumnuclear disk around the AGN ($R_{31} = 2.7$) and the starburst ring ($R_{31}=1.2$). Furthermore, they observed higher CO line ratios in more active star forming regions compared to less active ones ($R_{31} = 0.7-1.0$). Their analysis suggested that the elevated CO line ratio in the circumnuclear disk is a result of the AGN-driven outflow. The CO line ratio in IC~5063 was extensively studied by \citet{Dasyra2016} and \citet{Oosterloo2017}. They identified excited molecular gas along the radio jet and noted a higher CO line ratio in the central region compared to the outer region. For instance, \citet{Dasyra2016} conducted a calculation of the line ratio between CO (4--3) and CO (2--1), revealing a value of 12 in the robust jet-cloud interacting regions compared to an average of 5. \citet{Oosterloo2017} demonstrated that $R_{31} >1$ along the jets as opposed to 0.17 in the disk. Their study further indicated that the most elevated CO line ratio is proximate to the two radio lobes, implicating the shock generated by the outflow as the causative factor \citep{Morganti2021}. More recently, \citet{Audibert2023} identified a high CO line ratio between J$=3-2$ and J$=2-1$, specifically with $T_{CO (3-2)}/T_{CO (2-1)}=0.8$ in the region perpendicular to the radio jet, in contrast to the value of 0.4 in other regions within the type-2 QSO Teacup galaxy. Their findings underscored that this augmented CO line ratio is a consequence of the jet-driven shock interacting with the molecular gas.

Our observations align with the results reported by \citet{Garcia-Burillo2014}, where the $R_{31}$ values stand at 2.7 in the circumnuclear disk, 1.2 in starburst regions, and 0.4$-$0.5 in the halo. Consequently, the regions exhibiting high $R_{31}$ values in our study can be attributed to shock heating induced by the BH outflow.

\subsubsection{The Western Star Forming Region}

\begin{figure}
    \centering
    \includegraphics[width=5.5in]{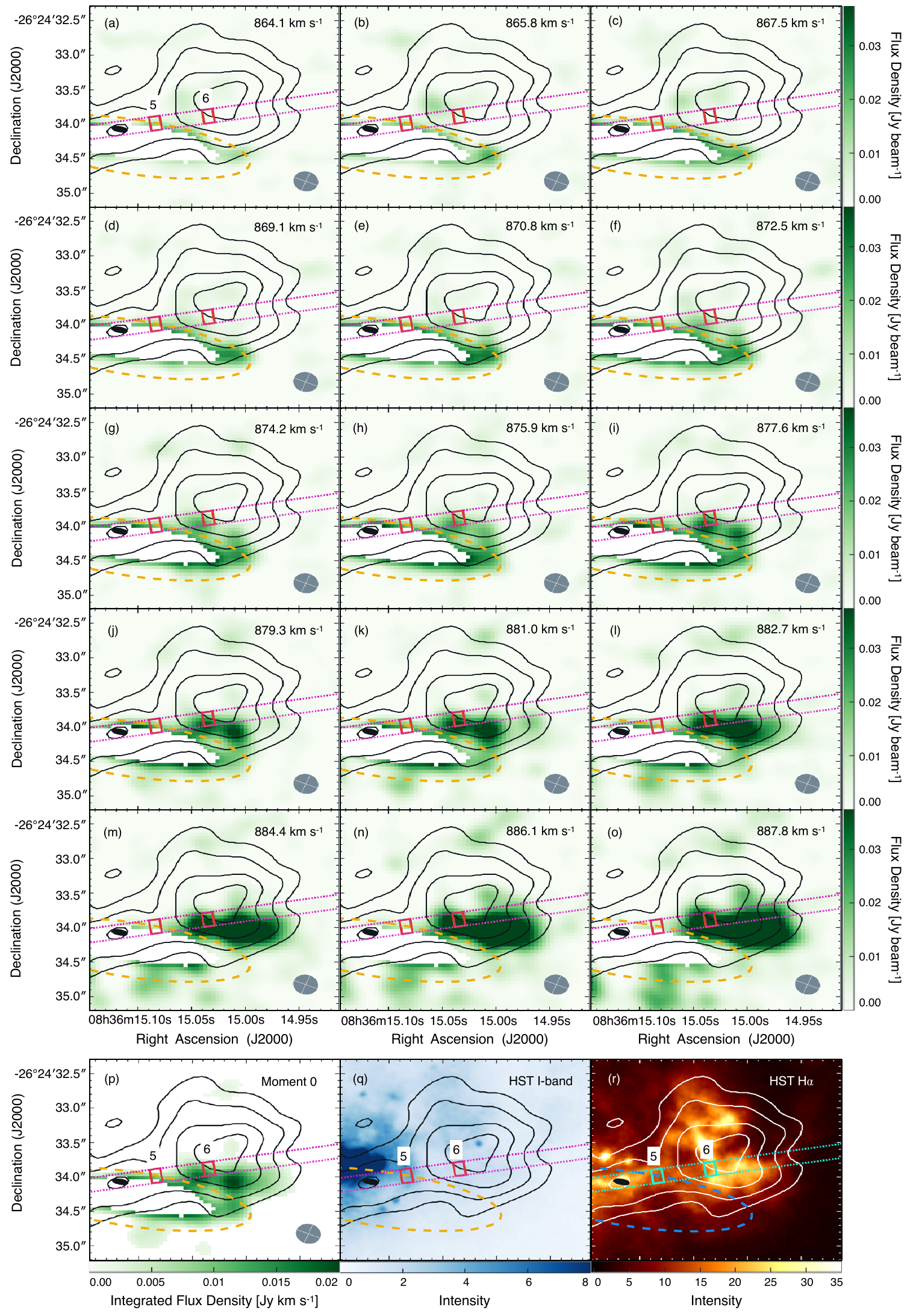}
    \caption{We present the channel maps of the CO (3--2) line within the velocity range of 864.1--887.8~km s$^{-1}$. In order to highlight the faint emission beyond the BH vicinity (indicated by the dashed orange line), we have masked the majority of CO gas. The 99~GHz continuum emission is depicted by black contours at levels of 50, 100, 200, and 300 $\mu$Jy beam$^{-1}$. The BH is represented by a solid black ellipse, corresponding to the central maximum of the 1.4~GHz continuum observed in \textit{LBA} \citep{Reines2012}. The diagonal dotted lines illustrate the spectroscopic slit used in the \textit{HST} observations, with a width of 0\farcs2. Additionally, the regions 5 (left) and 6 (right) previously discussed by \citet{Schutte2022} are depicted by red boxes. The ALMA beam size is provided in the bottom-right corner of each panel. The bottom row exhibits the total intensity map of the CO (3--2) line on the left, followed by the \textit{HST} I-band image in the middle, and the \textit{HST} H$\alpha$+continuum image on the right. In the H$\alpha$+continuum image (panel r), white contours represent the 99~GHz continuum map, cyan dotted lines depict the slit, and green boxes highlight regions 5 and 6 for improved visibility.}
\label{fig:co32_shock}
\end{figure}

While prominent at radio wavelengths (e.g., Figure \ref{fig:continuum}), the star-forming region to the west of the BH appears to suffer from significant extinction at optical wavelengths (e.g., Figures \ref{fig:co32_shock} and \ref{fig:co_ratio}). \citet{Schutte2022} estimate a stellar age of $<$3~Myr for region 6 (e.g., see Figure \ref{fig:co_ratio}) based on the equivalent width of the H$\alpha$ emission line. This is consistent with a scenario in which the star clusters have not had enough time to clear away their birth material, leading to high levels of extinction. We quantify their assertion with the molecular gas distribution.

Figure~\ref{fig:co32_shock} presents channel maps of the CO (3--2) line within velocities of 864--888~km s$^{-1}$ in panels (a) to (o), a total intensity map in panel (p), an \textit{HST} I-band image in panel (q), and an \textit{HST} H$\alpha$ plus continuum image in panel (r). The total intensity map was produced using 3D-Barolo \citep{Teodoro2015} with a primary S/N of 5 and a second S/N of 3 on the image cube shown in panels (a)--(o). In the channel maps and total intensity map, the bulk of the gas was masked to enhance contrast and highlight fainter emission. The red boxes (white boxes in panel r) in each panel of Figure~\ref{fig:co32_shock} indicate regions 5 and 6 defined by \citet{Schutte2022}, respectively. Figure~\ref{fig:co32_shock} demonstrates that there is a significant amount of molecular gas between regions 5 and 6, implying that the molecular gas is abundant enough to absorb the emission from young stellar populations embedded in the gas.

We estimate the optical extinction in this region as follows. The total intensity of the CO (1--0) line within the span of regions 5 and 6 is measured as $\rm W_{CO (1-0)} = 788.2 \pm 57.4$~K km s$^{-1}$, resulting in an estimated molecular gas column density of $\rm N(H_{2})= (2.21\pm 0.16)\times 10^{22} - (1.58\pm 0.12)\times 10^{23}$~cm$^{-2}$. This estimation is derived from the mean velocity-integrated flux density ratio of CO (3--2) to CO (1--0), represented as $\rm I_{CO (3-2)}/I_{CO (1-0)} =12.54$ in this region. The conversion factor $\rm X_{CO}= (0.28-2)\times 10^{20}$~cm$^{-2}$ (K km s$^{-1}$)$^{-1}$, corresponding to $\rm \alpha_{CO}= 0.6- 4.3$~$M_{\odot}$ (K km s$^{-1}$ pc$^{2}$)$^{-1}$, has been employed for this calculation. \citet{Kobulnicky1995} conducted observations of atomic hydrogen (\HI) using the VLA at D-, C-, and B-arrays. In this region, the total intensity of \HI\ was 1.52$\pm$0.05~Jy beam$^{-1}$ km s$^{-1}$, leading to a corresponding column density N(\HI) at (1.51$\pm$0.05)$\times$10$^{18}$~cm$^{-2}$, considering a synthesized beam size of 31\arcsec$\times$30\arcsec. Assuming a uniform distribution of \HI, the optical extinction can be computed using the relationship between N(H) and A$_{V}$ \citep{Zhu2017}, $\rm A_{V} = 4.81 \times 10^{-22} (N(HI) + 2 N(H_{2}))$. This calculation yields a range of $\rm A_{V} \sim 21-152$ mag. If the \HI\ is more concentrated in this region, these values would represent lower limits. On the other hands, the optical extinction is estimated to be $\rm A_{V} \sim 7-51$ for the eastern star forming region, supporting that the optically dark region is predominantly observed between region 5 and 6.

The 251 and 340~GHz images presented in Figure~\ref{fig:continuum} showed that the compact features of region 6 align with peaks in CO (3--2) and HCO$^{+}$ (3--2) shown in Figure~\ref{fig:moment_maps}. This observation is in line with the characteristics of bright and compact embedded clusters seen in other young clusters, e.g., NGC 253 \citep{Ando2017,Leroy2018,Levy2021,Mills2021} and NGC~4945 \citep{Emig2020}. To quantify our findings, we focused on the radio continuum blob located in the western star-forming region, corresponding to region 6 as defined in \citet{Schutte2022}. Dust and gas mass estimations were carried out within this region. For the calculation of dust mass, we utilized the 340~GHz flux density and employed the equation: 
\begin{equation}
    M_{dust} = {{S_{\nu} D^{2}} \over {\kappa_{\nu} B_{\nu} (T_{d})}}
\end{equation}
\citep{Launhardt1997}. We adopted a dust absorption coefficient of $\kappa_{\nu} = 1.3$~cm$^{2}$ g$^{-1}$ at 340~GHz \citep{Costa2021} and the dust temperature of 29~K \citep{Santangelo2009}. The measured flux density at 340~GHz is $S_{340GHz} = 610.6 \pm 30.0$~$\mu$Jy, resulting in a calculated dust mass of $(2.37\pm0.12)\times10^{3}$~M$_{\odot}$. The associated molecular mass is $M_{\rm H_{2}}= (7.79\pm0.15)\times 10^{4}-(5.65\pm0.11)\times 10^{5}$~M$_{odot}$ depending on the $\alpha_{CO}$. Consequently, the dust to molecular gas mass ratio falls within the range of 1:33--238, which is consistent with the typical value of Milky Way and other starburst galaxies, 1:120 \citep{Wilson2008}.

The existence of abundant molecular gas and high optical extinction value in the optically dark regions (regions 5 and 6) indicate that the molecular gas in these regions is compressed. In He 2--10, the shock generated by the BH outflow emerges as the most plausible source for the compression of molecular gas, contributing to the formation of embedded star clusters. This is an example of the ``positive feedback'' of the BH outflow.

The asymmetric distributions of molecular gas and embedded star formation provides a key to understanding the evolution of He~2--10. Given that He 2--10 has experienced a tidal interaction \citep{Kobulnicky1995}, the asymmetric distributions of molecular gas result from the tidal interaction by disturbing the galaxy potential. This asymmetric distribution of molecular gas might introduce the asymmetric embedded star formations via the asymmetric gas inflow, similar to the Central Molecular Zone of the Milky Way \citep{Battersby2020, Henshaw2022}.

\section{Conclusions \label{sec:conclusions}}

We re-imaged the ALMA archival data of He 2--10 using newly developed algorithms and studied properties of the continuum and molecular gas, including CO (1--0), CO (3--2), HCN (1--0), HCO$^{+}$ (1--0), and HCO$^{+}$ (3--2), focusing on the massive BH and its vicinity. We re-analyzed the radio--sub-mm SED of the BH with updated flux densities and investigated the kinematics and morphology of molecular gas around the BH. Our main results are summarized below.

\begin{itemize}
    \item The radio--sub-mm SED of the BH from $1.4-340$ GHz is dominated by synchrotron emission with a best-fit model of $S_\nu= (1349 \pm 217) \nu^{-0.52 \pm 0.06}$~$\mu$Jy (Figure~\ref{fig:BH_sed}).
    \item We find an elongated molecular gas structure ($\approx 128~{\rm pc} \times 34~{\rm pc}$) in the vicinity of the BH with a distinct velocity component in CO (1--0), CO (3--2), and HCO$^{+}$ (3--2) from the surrounding regions (Figure \ref{fig:BH_mom}). The mass of the molecular gas structure in the vicinity of the BH is $\rm M_{H_{2}} = (0.21-1.53) \times 10^{6} \; M_{\odot}$ for a range of metallicity-dependent values of $\alpha_{CO}$.  While the position of the BH is within this region, it is significantly offset from the peak intensity, which could explain why the BH is accreting at a low rate.
    \item  The ratio of CO (3--2) to CO (1--0) shows that the CO gas is highly excited in the BH vicinity (Figure \ref{fig:co_ratio}). The highest excitation values are located at the interface between the BH outflow and eastern region of triggered star formation identified by \citet{Schutte2022}. Our results suggest that the molecular gas is being compressed by the shock produced by the outflow from the BH, providing additional evidence for positive BH feedback.     
    \item The high excitation of the CO line ratio is also detected to the west of the BH, suggesting the other side of the bipolar flow is shocking dense molecular clouds in the western star forming region as well. We find high molecular gas column densities (${\rm N(H_2) \sim 10^{21-22}}~{\rm cm}^{-2}$) in this region, as well as high levels of extinction ($A_V \gtrsim 2-11$ mag). This is consistent with a scenario in which the infant star clusters with stellar ages $\lesssim$ 3 Myr \citep[][]{Schutte2022} have not yet had enough time to destroy or disperse the dense molecular clouds from which they formed.

\end{itemize}

The study presented in this paper constitutes the first investigation of the molecular gas properties in the immediate vicinity of the massive BH in He 2--10. While our results provide additional support for positive BH feedback, future observations with higher angular resolution are needed to resolve the sphere of influence of the BH and trace any potential molecular gas outflow directly. Furthermore, analyzing the higher transitions of the CO molecular gas and applying models for the photodissociation region (PDR) and/or X-ray dominated region (XDR) will provide additional information on physical parameters such as density, temperature, and outflow power, ultimately yielding more insight into the impact of BH feedback on nearby molecular gas.

\vspace{5mm}
\facilities{ALMA, VLA, LBA, HST}

\software{CASA \citep{CASA2022}, AIPS \citep{Greisen2003}, CARTA \citep{CARTA}, R \citep{Rsoftware}, 3D-Barolo \citep{Teodoro2015}}

\section*{Acknowledgement}
The authors express their gratitude to the anonymous reviewer for their valuable contributions in enhancing the quality of this paper. We are grateful to John Hibbard, Loreto Barcos-Munoz, Mark Lacy, Allison Costa, George Privon at NRAO, Seth Kimbrell, and Sheyda Salehirad at Montana State University for valuable discussions. A.E.R. acknowledges support for this work provided by NASA through EPSCoR grant number 80NSSC20M0231. This paper makes use of the following ALMA data: ADS / JAO.ALMA \#2011.0.00348.S, ADS / JAO.ALMA \#2012.1.00413.S, ADS / JAO.ALMA \#2015.1.01569.S, ADS / JAO.ALMA \#2016.1.00027.S, ADS / JAO.ALMA \#2016.1.00492.S, ADS / JAO.ALMA \#2019.1.01641.S. ALMA is a partnership of ESO (representing its member states), NSF (USA) and NINS (Japan), together with NRC (Canada), MOST and ASIAA (Taiwan), and KASI (Republic of Korea), in cooperation with the Republic of Chile. The Joint ALMA Observatory is operated by ESO, AUI/NRAO and NAOJ. This research has made use of NASA's Astrophysics Data System Bibliographic Services.

\bibliographystyle{aasjournal}

\end{document}